\documentclass[letter]{aa} 
\usepackage{graphicx}
\usepackage[dvips]{color}
\usepackage{txfonts}

\begin{document}

  \title{WISE/2MASS-SDSS brown dwarfs candidates using Virtual Observatory tools}


  \author{M. Aberasturi \inst{1,2}
\and
           E. Solano \inst{1,2}
           \and
           E. L. Mart\'{\i}n \inst{1}
          }

   \institute{Centro de Astrobiolog\'{\i}a (INTA-CSIC), Departamento de Astrof\'{\i}sica, P.O. Box 78, E-28691 Villanueva de la Ca\~{n}ada, Madrid, Spain\
              \email{miriam@cab.inta-csic.es, esm@cab.inta-csic.es, ege@cab.inta-csic.es}
        \and
             Spanish Virtual Observatory\\
%
             }

   \date{Received August 1, 2011; accepted XX xx, xx}

 
  \abstract
   {Massive far-red and infrared imaging  surveys in different bandpasses are the main contributors to the discovery of brown dwarfs (BDs). The Virtual
Observatory (VO) represents an adequate framework to handle these vast datasets efficiently and filter them out according to specific requirements. A statistically significant number of BDs is mandatory for understanding their general properties better for identifing peculiar objects. WISE, an all-sky survey in the mid-infrared, provides an excellent opportunity to increase the number of BDs significantly, in particular those at the lower end of the temperature scale. } 
   {We aim to demonstrate that VO tools are efficient in identifing and characterizing BDs by cross-correlating public catalogues released by large surveys.}
   {Using VO tools we performed a cross-match of the WISE Preliminary Release, the 2MASS Point Source and the SDSS Data Release 7 catalogues over the whole area of sky that they have in common ($\sim$ 4000 deg$^{2}$). Photometric and proper motion criteria were used to obtain a list of BD candidates. A temperature estimate is provided for each candidate based on their spectral energy distribution using VOSA, a VO tool for SED (Spectral Energy Distribution) fitting. We derive the spectral types from the effective temperatures. Distances, calculated from  the absolute magnitude- spectral type relation, place our candidates at 14-80 pc from the Sun, assuming that they are single.}
   {We have identified 31 BD candidates, 25 of which have already been reported in the literature. The remaining six candidates have been classified as L- (four) and T-type (two) objects. The high rate of recovery of known BDs ($\sim$ 90\% of the T dwarfs catalogued in 2MASS) demonstrates the validity of our strategy to identify them with VO tools. An application of this method for a deeper search that covers the whole sky in common to WISE and UKIDSS will be presented in a forthcoming work.}
  {}

   \keywords{Brown dwarfs --
                surveys --
                proper motions --
                astronomical databases miscellaneous -- Virtual Observatory tools
               }

   \maketitle
%

\section{Introduction}

Brown dwarfs (BDs), self-gravitating objects that form like stars but do not get enough mass to maintain a sufficiently high temperature and pressure in their cores for stable hydrogen 
fusion, provide a natural link between very low-mass stars and gaseous giant planets. Theoretically proposed fifty years ago (\cite{1963ApJ...137.1121K}), it was not until 
1995 when the first BDs were discovered (\cite{1995Natur.377..129R}; \cite{1995Natur.378..463N}). Since then, hundreds of BDs have been found mainly thanks to the advent of large-area optical and 
near-infrared surveys such as the Two Micron All Sky Survey (2MASS, \cite{2006AJ....131.1163S}; \cite{2007AJ....134.1162L}), the DEep Near Infrared Survey 
of the Southern Sky (DENIS, \cite{1997Msngr..87...27E}; \cite{1997A&A...327L..25D}; \cite{1999AJ....118.2466M}), the Sloan Digital Sky Survey 
(SDSS, \cite{2000AJ....120.1579Y}; \cite{2006AJ....131.2722C}), the UKIRT Infrared Deep Survey (UKIDSS,  \cite{2007MNRAS.379.1599L}; 
\cite{2007MNRAS.379.1423L}; \cite{2010MNRAS.406.1885B}) and the Canada-France Brown Dwarf Survey (CFBDS, \cite{2008A&A...484..469D}; \cite{2011AJ....141..203A})\footnote{See http://dwarfarchives.org for an updated list of L and T dwarfs}.

The Wide-field Infrared Survey Explorer (WISE; \cite{2010AJ....140.1868W}) is a NASA mission that has mapped the sky at 3.4 (W1), 4.6 (W2), 12 (W3), and 22 (W4) 
$\mu$m in 2010 with an angular resolution of 6.1", 6.4", 6.5", and 12.0", respectively. WISE achieved 5$\sigma$ point-source sensitivities better than 
0.08, 0.11, 1 and 6 mJy in unconfused regions on the ecliptic in the four bands. The dataset obtained by the WISE imaging survey constitutes an excellent resource for finding new brown dwarfs, in particular the coldest members, because they emit a substantial part of the spectral flux in the mid-infrared. The recent discoveries of T dwarfs reported by different authors 
(Burgasser et al. 2011; Mainzer et al. 2011; \cite{2011A&A...532L...5S}) confirm the high expectations placed on the mission. The WISE Preliminary Release\footnote{http://irsa.ipac.caltech.edu/Missions/wise.html} is available to the astronomical community since April 14, 2011 and includes photometric information for over 257 million objects observed during the first 105 days of the survey. 

The Virtual Observatory\footnote{http://www.ivoa.net} (VO) is an international initiative designed to help the astronomical community in the exploration
of the digital, multi-wavelength universe that is resides in the
astronomical data archives. The VO is already an operational research infrastructure, as demonstrated by the growing number of VO-papers published in the last years ($>$ 50 since 2009)\footnote{http://www.euro-vo.org/pub/fc/papers.html}. In this work we have made use of VO tools to benefit from an easy data access and analysis.

We present here the identification of 31 BDs (25 known and 6 strong candidates not previously reported in the literature) identified in the sky area in common to the WISE Preliminary Release and the 2MASS Point Source and SDSS Data Release 7 catalogues. In Sect. 2 we describe the methodology devised to search for T dwarfs. In Sect. 3 we derive proper motions, effective temperatures ($T_\mathrm{eff}$), spectral types, and distances for our candidates. Finally, we summarize conclusions in Sect. 4.


\section{Candidate selection}

\begin{figure*}[h!]
   \centering
   \includegraphics[width=19cm]{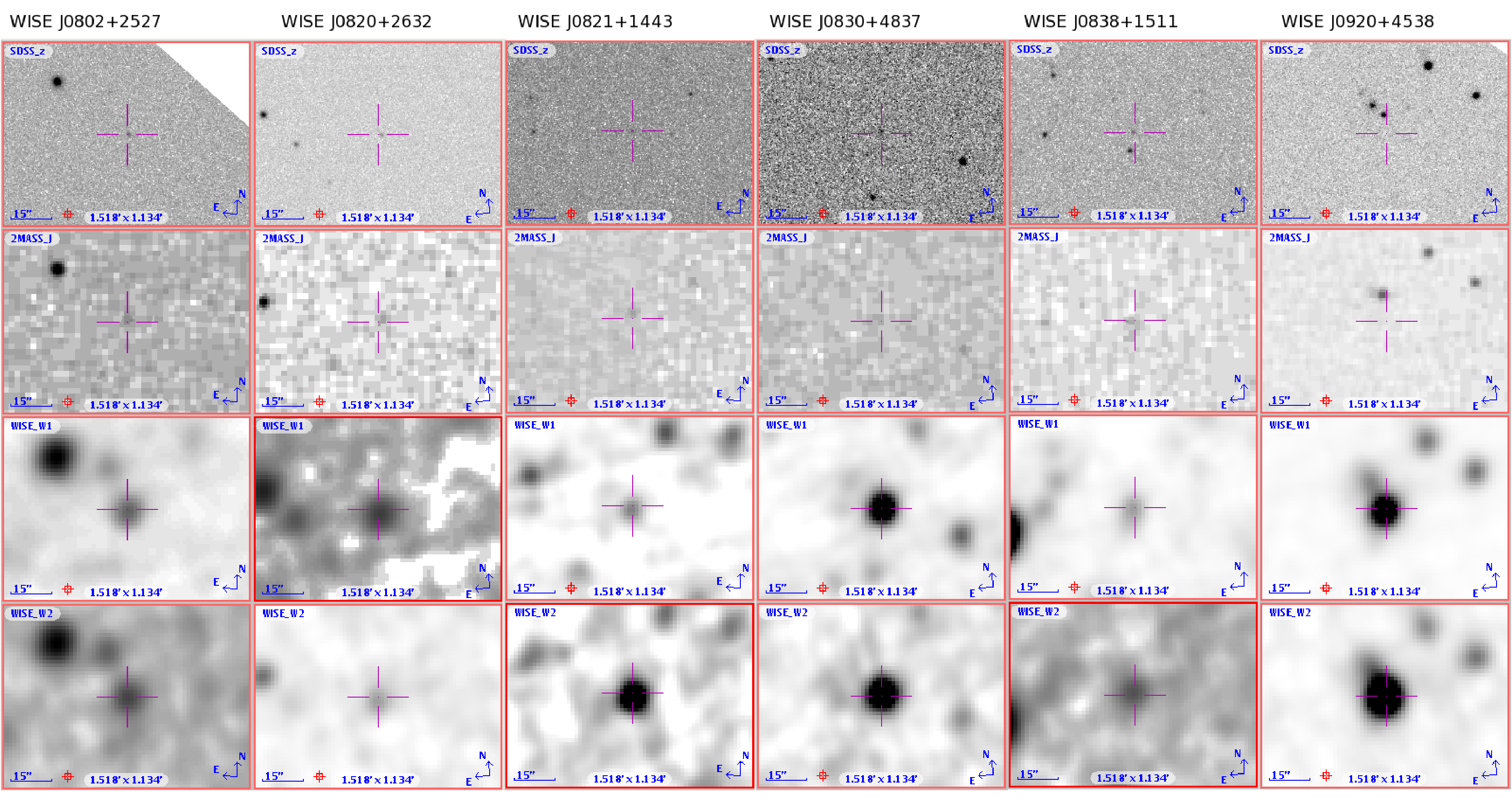}
   \caption{Four-band charts showing the area around the six new BD candidates. The fields are 1.5\arcmin $\times$ 1.1\arcmin oriented with North up and East to the left. For each source, the SDSS(\textit{z'}), 2MASS(\textit{J}), WISE(W1) images are centered on the WISE(W2) coordinates. Fast motion between 2MASS and WISE epochs is clearly visible in WISE J0920+4538.} 

   \label{secuencia}
\end{figure*}
\begin{figure}[t!]
   \centering
   \includegraphics[width=7cm]{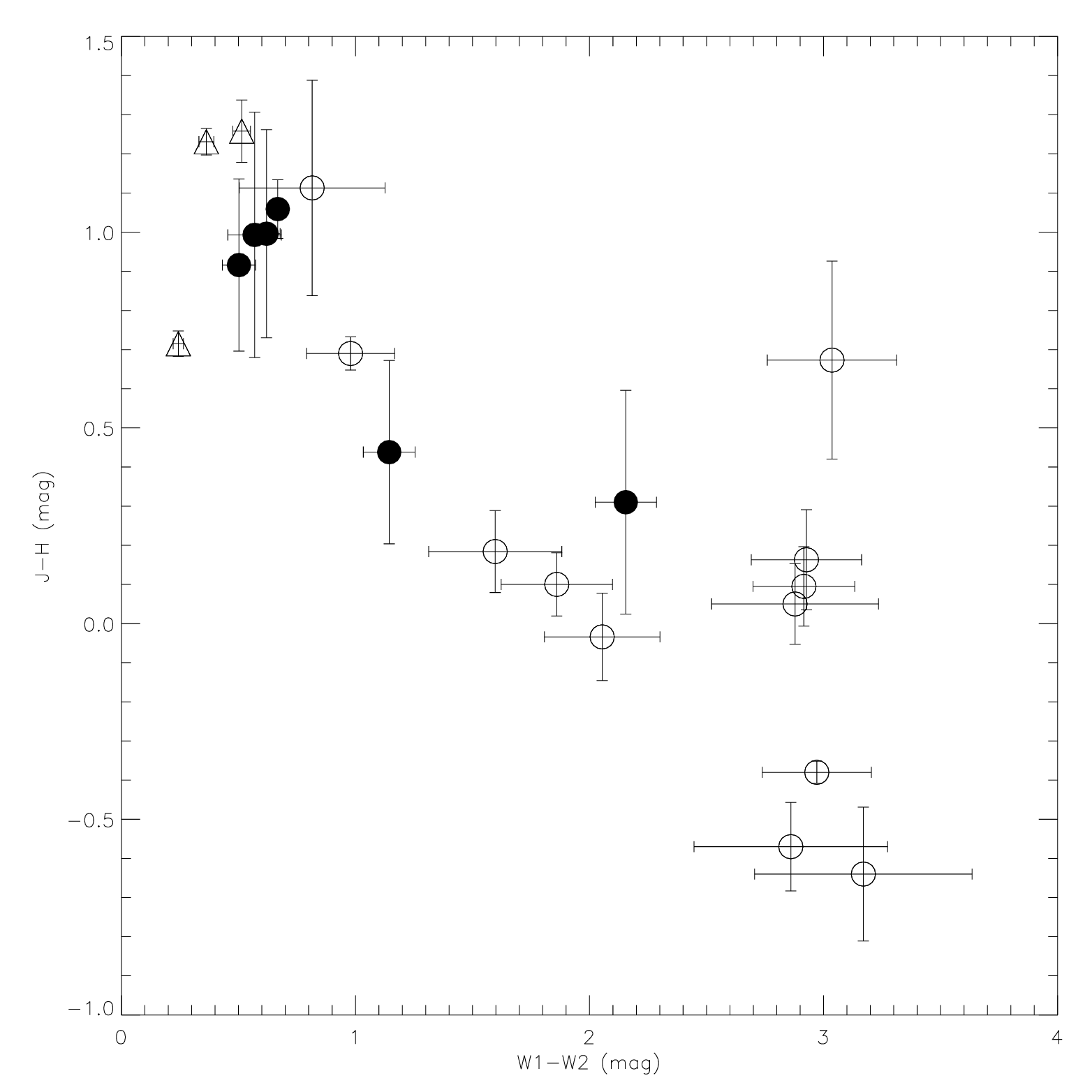}
   \caption{Colour-colour diagram of our six candidates (black full circles) compared with a sample of known L  and T dwarfs (triangles and circles, 
respectively) observed with WISE (Mainzer et al. 2011; \cite{2011A&A...532L...5S}; Burgasser et al. 2011.)} 
   \label{diagrama}
\end{figure}

We built a VO-workflow with the STILTS\footnote{http://www.star.bris.ac.uk/$\sim$mbt/stilts/} scripting capabilities  with  criteria that must accomplish our preliminary list of candidates. The workflow consisted in the following steps: 

\begin{itemize}

\item Cross-match. To avoid memory overflow problems associated to the data processing of large volumes of data, we divided the common WISE-2MASS-SDSS sky into overlapping circular regions of 30\arcmin. After different tests, we adopted a matching radius of 20\arcsec$ $ to ensure that objects with high proper motion were not left out while at the same time avoiding an unmanageable number of false positives. Only the closest 2MASS and SDSS counterparts to each WISE source were considered. \\

\item Filters on photometry. \\

\begin{itemize}

\item[$\bullet$] Xflg=0, Aflg=0, to avoid sources flagged in 2MASS as minor planets or contaminated by nearby extended sources.
\item[$\bullet$] 9 $<$ \textit{J}, to exclude of too bright sources that typically have associated large uncertainties. 
\item[$\bullet$] Qflg(\textbf{\textit{J}}) $\neq$ "U", to discard sources with upper limits in the \textit{J} band.
\item[$\bullet$] \textit{g'} $>$ 22.2, \textit{r'} $>$ 22.2 (SDSS limiting magnitudes), because T dwarfs should not be detected in these bands.
\item[$\bullet$] \textit{i'} $>$ 21.3 (SDSS limiting magnitude) or (\textit{i'}-\textit{z'}) $\geq$ 3 (Hawley et al. 2002). 
\item[$\bullet$] \textit{z'} $<$ 20.5 (SDSS  limiting magnitude) 
\item [$\bullet$] (W1-W2)$>$0.5 , W2$<$15.5, (W2-W3)$<$ 2.5. The W1 and W2 WISE bands were specifically designed to distinguish T dwarfs from background 
sources. W1 includes much of the CH$_{4}$ fundamental absorption band and W2 includes the pseudo-continuum peak that is present for all  objects cooler than 3000 K (\cite{2003ApJ...596..587B}; \cite{2004AJ....127.3516G}). \cite{2011ApJ...726...30M}  suggest a (W1-W2) $>$ 1.8 for cool T dwarfs. Because we search for T dwarfs at all spectral subtypes, we relaxed this criterion up to the stellar boundary, (W1-W2)$>$ 0.5. A side effect of this approach was the identification of four new L-type candidates. The (W2-W3)$<$ 2.5 criterion was selected to exclude extragalactic sources (\cite{2010AJ....140.1868W}). W2=15.5 indicates the limiting magnitude (5$\sigma$) in the W2 band.
\item[$\bullet$] (\textit{J}-W2)$>$1.8 and (\textit{z'}-\textit{J})$>$2.5 as suggested by \cite{2011ApJ...726...30M} and \cite{2002AJ....123.3409H}, respectively.\\
\end{itemize} 

\item Filter on astrometry and proper motions (PMs). \\

\begin{itemize}
\item[$\bullet$] The angular separation between the WISE source and the 2MASS counterpart must be greater than 1\arcsec$ $ (WISE 2$\sigma$ astrometric accuracy). Assuming a minimum difference epoch between 2MASS and WISE of nine years (Feb 2001 - Jan 2010), this criterion implies that we are loosing objects with PMs lower than 0.11\arcsec/yr. On the other hand, the 20\arcsec $ $ matching radius will limit the detection of T dwarfs with PM higher than 1.53\arcsec/yr $ $given the maximum difference epoch between 2MASS and WISE (Jun 1997 - Apr 2010). 
\item[$\bullet$]There must not be counterparts in the source table of the Supercosmos Science Archive (\cite{2001MNRAS.326.1279H}) at less than 1\arcsec$ $ from the WISE source.
\item[$\bullet$] The differences in the PMs derived from WISE-2MASS and WISE-SDSS must be less than 50\%.
\end{itemize}
\end{itemize}
This query returned 138 candidates. They were visually inspected using the scripting capabilities of Aladin\footnote{http://aladin.u-strasbg.fr/} (\cite{2000A&AS..143...33B}). Aladin is a VO-compliant software that allows users to visualize and analyze digitised astronomical images, and superimpose entries from astronomical catalogues or databases available from VO services. Sources from DENIS, 2MASS, SDSS, Supercosmos and UKIDSS as well as images from UKIDSS, 2MASS and SDSS were used in the analysis. One hundred and seven candidates were rejected due to several reasons (artifacts, presence of a nearby bright star, etc.), the most likely being the mismatch between WISE sources and 2MASS/SDSS counterparts: due to the different depth of the surveys it may happen that a 2MASS source is wrongly associated to a SDSS source without a real counterpart in 2MASS. Thirty one candidates passed the visual inspection. Twenty three sources were identified as known BDs (7 L- and 16 T-dwarfs) in a cross-match with SIMBAD\footnote{http://simbad.u-strasbg.fr/simbad/} and the DwarfArchives, providing an independent sanity check of our selection method. Two out of the remaining eight 
candidates were published during the preparation of this letter: WISE J1625+1528 (\cite{2011arXiv1106.3105D}) and WISE J1627+3255 
(\cite{2011AJ....142...57G}). Multiwavelength charts and the astrometric and photometric information of the six newly discovered BD candidates are given 
in Fig.~\ref{secuencia} and Table~\ref{tabla}, respectively. Figure ~\ref{diagrama} compares in a color-color diagram our six candidates with a sample of 
known L and T dwarfs (\cite{2011ApJ...726...30M}, \cite{2011A&A...532L...5S}, \cite{2011ApJ...735..116B}). All our candidates have  W1,W2 signal-to-noise ratios (SNR) $\geq$ 10 and can be considered as point 
sources (w2rchi2 $\leq$ 3). They were also searched for common PM companions within a radius of 10\arcmin$ $ in the WISE and 2MASS catalogues but none was found.

Finally, we assessed the efficiency of our search by estimating the false negative rate (number of known T dwarfs that were not rediscovered in the analysis). Thirty six T dwarfs catalogued in the DwarfArchives were not found with our methodology. Most of them (33) were too faint to be detected in 2MASS and the remaining three did not meet some of the photometric criteria explained above.

\begin{table*}
\centering
\caption{Astrometry, photometry and physical parameters for the six new BDs candidates. See Sect. 3 for a detailed description on how effective temperatures, spectral types, absolute magnitudes and distances were obtained. \textit{J} and W2 absolute magnitudes were only computed for L and T dwarfs respectively. Also, the uncertainty in W3 is not provided for four of the candidates. According to the WISE source catalogue user's guide, this happens if the W3 profile-fit magnitude is a 95\% confidence upper limit or if the source is not measurable.}
\label{tabla}
\scriptsize

\hspace{-1cm}
\begin{tabular}{ccccccc}
\noalign{\smallskip}
\hline
\noalign{\smallskip}
Parameter		& WISE J0802+2527	& WISEJ0820+2632   &  WISE J0821+1443  	      & WISE J0830+4837   & WISE J0838+1511 & WISE J0920+4538 \\
\noalign{\smallskip}
\hline
\noalign{\smallskip}
RA(J2000)        	& 08 02 02.86 & 08 20 07.35 & 08 21 31.63 & 08 30 41.67 & 08 38 11.45 &  09 20 55.41  \\
Dec(J2000)       	&$+$25 27 18.5&$+$26 32 28.5&$+$14 43 19.4&$+$48 37 15.0&$+$15 11 15.1&$+$45 38 56.4    \\
$\mu_{\alpha}cos\delta$ ($\arcsec$/year)&   -0.011$\pm$0.026       &  0.087$\pm$0.042             &  -0.115$\pm$0.035    &  -0.125$\pm$0.026   &   -0.121$\pm$0.031     &  -0.075 $\pm$0.010      \\
$\mu_{\delta}$ ($\arcsec$/year)  	&  -0.125$\pm$0.055        &  -0.208$\pm$0.062             &  -0.297$\pm$0.054  &  -0.058$\pm$0.050    &   -0.032$\pm$0.051    &  -0.833$\pm$0.045       \\
\textit{z'}	         & 19.238$\pm$0.069 & 19.514$\pm$0.067 & 19.732$\pm$0.118 & 19.905$\pm$0.153 &19.907$\pm$0.142 &21.002$\pm$0.658 \\
Y (UKIDSS)               & 17.914$\pm$0.012 & 18.145$\pm$0.029 &                  &                  &                 &                 \\
\textit{J} (2MASS) 	 & 16.713$\pm$0.157 & 16.988$\pm$0.222 & 16.825$\pm$0.155 & 17.275$\pm$0.199 &16.645$\pm$0.159 &15.223$\pm$0.052 \\
\textit{J} (UKIDSS)      & 16.626$\pm$0.010 & 17.036$\pm$0.015 &                  &                  &                 &                  \\
\textit{H} (2MASS)       & 15.797$\pm$0.154 & 15.995$\pm$0.221 & 16.515$\pm$0.240 & 16.279$\pm$0.176 &16.207$\pm$0.172 &14.164$\pm$0.054 \\      
\textit{H} (UKIDSS)      & 15.937$\pm$0.019 & 16.487$\pm$0.038 &                  &                  &                 &                  \\
\textit{K$_{s}$} (2MASS) & 15.397$\pm$0.151 & 15.881$\pm$0.251 &                  & 15.596$\pm$0.165 &                 &13.728$\pm$0.050 \\
\textit{K$_{s}$} (UKIDSS)& 15.248$\pm$0.016 & 15.883$\pm$0.028 &                  &                  &                 &                  \\
W1                       & 14.758$\pm$0.037 & 15.578$\pm$0.057 & 16.438$\pm$0.114 & 14.620$\pm$0.034 & 15.712$\pm$0.068&13.059$\pm$0.025 \\
W2                       & 14.256$\pm$0.060 & 15.009$\pm$0.099 & 14.283$\pm$0.064 & 14.000$\pm$0.047 & 14.568$\pm$0.087&12.391$\pm$0.026  \\
W3                       & 11.965$\pm$---   & 12.658$\pm$---   & 12.572$\pm$---   & 12.828$\pm$0.539 & 12.285$\pm$---  &11.063$\pm$0.115 \\
W1$-$W2 & 0.502 &0.569 & 2.155 & 0.62 & 1.144 & 0.668 \\
W2$-$W3 & 2.291 & 2.351& 1.711 & 1.172 & 2.283 & 1.328\\

$T_\mathrm{eff}$  (VOSA) (K)& 1800 & 2000 & 700 & 1700 & 900 & 1700 \\
SpT ($T_\mathrm{eff}$) & L3-L4 & L2-L3 & T8 & L4-L5 & T6-T7 & L4-L5 \\
$M_{J}$   & 13.08 & 12.48 &        & 13.31 &         & 13.31        \\
$M_{W2}$           &       &       & 13.47  &       &  13.07  &        \\
d (pc)                                 & 53 & 80 & 15  & 62 &  20  &  24\\
\noalign{\smallskip}
\hline
\noalign{\smallskip}
\noalign{\smallskip}
\hline
\noalign{\smallskip}
\end{tabular}

\end{table*} 
\begin{figure*}[h!]
    \label{VOSA}
    \centering
    \includegraphics[width=11cm]{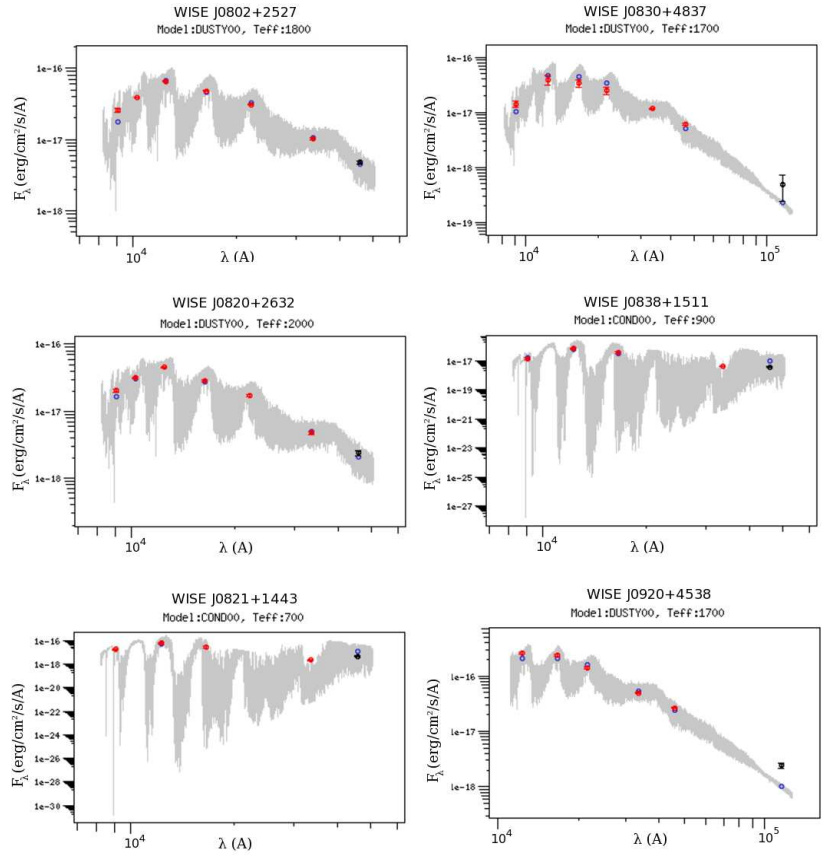}
    \caption{VOSA SED fitting for our six BD candidates. Catalogue and synthetic photometric points are represented in red and blue, respectively. Points 
not considered in the fitting are shown in black. Spectral bands are those listed in Table 1. The theoretical spectrum that best fits (and from which the synthetic photometry was computed by 
convolving with the corresponding filter profiles) is also plotted in grey.}
 \end{figure*}     

\section{Physical parameters}

\subsection{Proper motions}

The long baseline between WISE and 2MASS/SDSS, the excellent astrometry of these catalogues (less than 0.5\arcsec$ $ with respect to 2MASS in the case of WISE (\cite{2010AJ....140.1868W})), together with the fact that brown dwarfs are nearby objects and therefore typically show high proper motions facilitates the measurement of reliable angular separations between the WISE sources and their respective counterparts in the 2MASS and SDSS catalogues. 

Proper motions were computed using a linear least-squares fit to the coordinates given in the WISE, 2MASS and SDSS catalogues and weighted by their astrometric errors. No correction for parallactic motion was considered. The small separation between the 2MASS and SDSS positions (less than 1$\arcsec$ for five of our candidate BDs) and the lack of 2MASS $K_{s}$ photometry for two of them are the main reasons for their omission in previous 2MASS/SDSS-based PM searches.

\subsection{Effective temperature and spectral types}
Effective temperatures for our candidate BDs were obtained from the spectral energy distribution (SED) $\chi^{2}$ fitting between the observed photometry and a suite of collections of theoretical models. The temperature determination was carried out using VOSA\footnote{http://svo.cab.inta-csic.es/theory/vosa/} (Virtual Observatory SED Analyzer, \cite{2008A&A...492..277B}). VOSA is a VO-tool designed to query several photometric catalogues that are accessible through VO services as well as VO-compliant theoretical models and perform a statistical test to determine which model reproduces the observed data best. $T_\mathrm{eff}$ is then estimated from the best fit. Two different model collections were used in our analysis: DUSTY (\cite{2001ApJ...556..357A}) and COND (\cite{2000ApJ...542..464C}), which have $T_\mathrm{eff}$ models ranging from 4000 to 100K and 3900 to 500K in 100K steps, respectively. Effective temperatures obtained with VOSA are shown in Fig. \ref{VOSA} and Table \ref{tabla}.
 
Spectral types were then estimated following the effective temperature--spectral type relation given in \cite{2005ARA&A..43..195K}. The 23 BDs found in our analysis  with spectral types previously reported in the literature were used as benchmarks to calibrate the method. Typical errors of $\sim$0.5 spectral types were found. The main contribution to the error budget comes from the degeneracy in the $T_\mathrm{eff}$-- spectral. type relation between L4 and T4.

\subsection{Distances}

Distances for the L dwarfs were calculated using the absolute \textit{J} magnitude (2MASS) -- spectral type relation given in 
\cite{2003AJ....126.2421C}, whereas the distances to the T dwarfs were estimated using the absolute W2 magnitude (WISE) -- spectral type relation derived 
by \cite{2011ApJ...735..116B}. The calculated values for the distances are given in Table \ref{tabla}. The uncertainties in the derived distances are dominated by the uncertainty in the spectral type because the errors in the 2MASS J photometry are typically only 0.1-0.2 mag, whereas the errors in the spectral type are 0.5 types. This leads to a 30\% uncertainty on distance, assuming all our candidates are single. An unresolved multiplicity would imply an underestimate of distance.

\section{Conclusions}

Taking advantage of VO tools, we reported the discovery of six BD candidates in the region of the sky in common to 2MASS Point Source, SDSS (Data Release 7) and the WISE Preliminary Release catalogues. The six candidates are clearly visible in the WISE W1 and W2 bands with SNR $\geq$ 10 and show physical parameters typical of L and T-type objects. The number of new candidates is remarkable, considering that 2MASS has been extensively searched for ultracool dwarfs. 

These results clearly show how new surveys and the use of VO tools can help to mine older surveys: all but one of our BD candidates are very close to the 2MASS limiting magnitude, which puts them beyond the limits of previous, much shallower 2MASS-based BD searches (e.g. J$<$16, \cite{2007AJ....134.1162L}). Also remarkable is the fact that the methodology used in this paper is not limited to brown dwarfs but can be easily extrapolated to searches for other rare objects (e.g. high-z quasars).

Finally, our work clearly  demonstrates the suitability of exploiting WISE data following a VO methodology and increases the expectations of building an accurate census of substellar objects in the solar vicinity.

\begin{acknowledgements}
 This research has made use of the WISE data hosted at the NASA/ IPAC Infrared Science Archive, which is operated by the Jet Propulsion Laboratory, California Institute of Technology, under contract with the National Aeronautics and Space Administration. This publication has also made use of the SIMBAD, Vizier and Aladin services, operated at CDS, Strasbourg, France. Our research has benefitted from the M, L, and T dwarf compendium housed at DwarfArchives.org and maintained by Chris Gelino, Davy Kirkpatrick, and Adam Burgasser. Effective temperatures were estimated using VOSA, a VO-tool developed under the Spanish Virtual Observatory project supported from the Spanish MICINN through grant AyA2008-02156.
\end{acknowledgements}

\end{document}